\documentclass[aps,12pt,a4paper]{revtex4}

\usepackage{graphicx}
\usepackage{amsmath}
\usepackage{amssymb,amsbsy}
\usepackage{amsfonts}
\usepackage{wasysym}
\usepackage[export]{adjustbox}
\usepackage{xspace}
\usepackage[usenames]{color}
\usepackage{dcolumn}
\usepackage{bm}
\usepackage{mathrsfs}
\usepackage[colorlinks=true,urlcolor=blue]{hyperref}
\usepackage[all]{hypcap}
\usepackage[utf8]{inputenc}
\usepackage[T1]{fontenc}
\usepackage{multirow}

\usepackage{ulem}
\definecolor {darkgreen}{rgb}{0.2,0.7,0.2}

\newcommand{\IIm}{{\rm Im}}
\newcommand{\RRe}{{\rm Re}}

\newcommand{\eq}{\begin{equation}}
\newcommand{\be}{\begin{equation}}
\newcommand{\eeq}{\end{equation}}
\newcommand{\ee}{\end{equation}}

\newcommand{\mnras}{Mon.\ Not.\ Roy.\ Astron.\ Soc.\ }

\newcommand{\aap}{Astron.\ Astrophys.\ }

\newcommand{\SISSA}{SISSA, Via Bonomea 265, 34136 Trieste, Italy.}
\newcommand{\INFN}{INFN, sez. Trieste, Via Valerio 2, 34127 Trieste, Italy.}
\newcommand{\IFPU}{IFPU, Institute for Fundamental Physics of the Universe, Via Beirut 2, 34014 Trieste, Italy.}
\newcommand{\IAP}{Institut d'Astrophysique de Paris, CNRS \& Sorbonne Universit\'es, UMR 7095, 98 bis bd Arago, 75014 Paris, France.}

\begin{document}

\title{ Numerical investigation of plasma-driven superradiant instabilities}

\author{Alexandru Dima}
\email{adima@sissa.it}
\address{\SISSA}
\address{\INFN}
\address{\IFPU}
\author{Enrico Barausse}
\email{barausse@sissa.it}
\address{\SISSA}
\address{\INFN}
\address{\IFPU}
\address{\IAP}

\begin{abstract}
Photons propagating in a plasma acquire an effective mass $\mu$, which is given by the plasma frequency and which scales with
the square root of the plasma density. As noted previously in the literature, for electron number densities $n_e\sim 10^{-3}$ cm$^{-3}$
(such as those measured in the interstellar medium)
the effective mass induced by the plasma is $\mu \sim 10^{-12}$ eV.
This would cause superradiant instabilities for spinning black holes of a few
tens of solar masses.
An obvious problem with this picture is that densities in the vicinity of black holes
are much higher than in the interstellar medium because of accretion, and possibly also pair production.
We have conducted numerical simulations of the superradiant instability in spinning black holes surrounded by a plasma with
density increasing closer to the black hole, in order to mimic the effect of accretion.
While we confirm that superradiant instabilities appear for plasma densities that are
sufficiently low near the black hole, we find that
astrophysically realistic accretion disks are unlikely to trigger such instabilities.
\end{abstract}

\maketitle

\date{\today \hspace{0.2truecm}}

\section{Introduction}
 The detection of gravitational waves~\cite{Abbott:2016izl} by Advanced LIGO~\cite{TheLIGOScientific:2014jea} and Advanced Virgo~\cite{TheVirgo:2014hva}
was a major milestone in the history of astronomy. Not only have these observations
confirmed directly the existence of gravitational waves (already indirectly proven by binary pulsars~\cite{Hulse:1974eb,Taylor:1982zz}), but
they also provide a way to test astrophysical models for the formation of binaries of compact objects~\cite{LIGOScientific:2018jsj} and to
verify the validity of general relativity in the hitherto unexplored highly relativistic strong field regime~\cite{Abbott:2018lct,LIGOScientific:2019fpa}.
Crucial in both respects are the spins of the binary components, which could in principle be large, especially for black holes (BHs).

BH spins provide useful diagnostics to discriminate between astrophysical formation scenarios for binaries~\cite{TheLIGOScientific:2016htt,Zevin:2017evb}, e.g.
the field binary formation channel~\cite{1976Paczynski} vs the dynamical one~\cite{Benacquista2013}. Moreover, via a mechanism
known as \textit{superradiance}~\cite{Penrose:1969pc,Zeldovich1971,Starobinsky1973a,Starobinsky1973b,Detweiler:1980uk,Press:1972zz,Cardoso:2004nk,Brito:2015oca}, moderate to large BH spins
would allow for testing the presence of ultralight bosons and the existence of event horizons~\cite{Arvanitaki:2009fg,Arvanitaki:2010sy,Brito:2014wla,Brito:2017wnc,Brito:2017zvb}.
Superradiance occurs when BHs of sufficiently high spins are surrounded by light bosons with Compton wavelength comparable to the event horizon's size, or when a reflective
or partially reflective mirror is placed outside the event horizon to mimic possible deviations from the BH paradigm.
Akin to the Penrose process~\cite{Penrose:1969pc}, of which it constitutes the ``wave'' generalization, the superradiant instability proceeds to extract energy and angular momentum from the BH, transferring it to the light boson field or,
 in the mirror case, to perturbations of the metric. As a result, the BH spins down until the instability is quenched (which happens typically for dimensionless
spin parameters $\chi = cJ/(GM^2)\sim 0.1$--$0.4$, c.f.~e.g. Fig. 1 of Ref.~\cite{Brito:2017wnc}).

Unfortunately, LIGO and Virgo have so far gathered (mild) evidence for non-zero spins in only two  of the ten BH binaries detected so far~\cite{Abbott:2016nmj,LIGOScientific:2018mvr}. This is quite surprising,
since BHs in X-ray binaries have spins (measured by fitting the continuum spectrum~\cite{McClintock:2013vwa} or iron K$\alpha$ lines~\cite{Steiner:2012vq}) that seem distributed uniformly between zero and
the maximal Kerr limit~\cite{Middleton:2015osa}, and because field binary formation models tend to predict non-vanishing values for the effective spin parameter $\chi_{\rm eff}$ measured by LIGO/Virgo~\cite{Gerosa:2018wbw}. Since $\chi_{\rm eff}$
is only sensitive to the projections of the spins on the orbital angular momentum direction, it is of course possible that LIGO binaries may simply have moderate/large {\it and} randomly oriented spins. However, in the field binary scenario,
random spin orientations are generally produced only  by large supernova kicks~\cite{Farr:2011,Tauris:2017omb}, which
are disfavored by the merger rate measured by LIGO/Virgo~\cite{Abbott:2017vtc}. The dynamical formation channel, where random orientations are natural~\cite{Rodriguez:2016vmx}, typically predicts
fewer coalescences than observed.

An interesting proposal to explain the low values of the LIGO/Virgo BH spins was put forward by
Conlon and Herdeiro~\cite{Conlon:2017hhi} (see also Ref.~\cite{pani_loeb}), who noted that spinning BHs surrounded by a tenuous plasma
may be susceptible to superradiant instabilities. Indeed, the plasma induces a change in the dispersion
relation of photons propagating through it~\cite{Langmuir1929}:
$\omega^2=k^2+\omega_p^2\,,$
where  $n_e$ is the electron number density and $\alpha={e^2}/(4\pi)$ is
 the fine structure constant in natural units
($G=\hbar=c=1$, which we will adopt throughout this paper). As a result, photons acquire an effective mass equal to the plasma frequency, $\mu=\omega_{\rm p}$, defined by
\begin{equation}\label{eq:disprel}
\omega_{\rm p}=\sqrt{\frac{4\pi\alpha n_e}{m_e}}=1.2\cdot 10^{-12}\sqrt{\frac{n_e}{10^{-3}\text{cm}^{-3}}}\text{eV}.
\end{equation}
For $n_e\sim 10^{-3}$--$10^{-2}$ cm$^{-3}$, corresponding to
typical conditions in the interstellar medium (ISM)~\cite{Cordes:2002wz,Schnitzeler:2012,Yao_2017}, the photon develops an effective mass $\mu\sim 10^{-12}$--$10^{-10}$ eV, whose
wavelength is comparable to the gravitational radius of LIGO/Virgo BHs.
Indeed, for these densities the ``mass coupling'' -- i.e. the ratio between the BH's gravitational radius $M$ and the  Compton wavelength $1/\mu$ --
is
\begin{equation}\label{eq:masscoup}
\mu M = \left(\frac{M}{M_{\astrosun}} \right) \left( \frac{\mu}{10^{-10}\text{eV}}\right) \sim O(0.01)\text{--}O(1)\left(\frac{M}{M_{\astrosun}} \right) \,.
\end{equation}
Since the fastest growing superradiant modes are
found numerically for nearly extremal BHs ($\chi=0.99$) and $\mu M \sim 0.42$~\cite{Dolan:2007mj}, Ref.~\cite{Conlon:2017hhi} argues that
LIGO/Virgo BHs immersed in a plasma with
$n_e\sim 10^{-3}$--$10^{-2}$ cm$^{-3}$  are potentially unstable to superradiance, i.e. rotational energy can be extracted from them and transferred to a ``photon cloud''
surrounding the BH, and as a result the BH spin decreases.

As already pointed out in  Ref.~\cite{Conlon:2017hhi}, however, one obvious problem with this scenario is its applicability to accreting BHs in the real Universe.
 The standard picture of accretion onto BHs assumes that the accreting gas will generally have sizeable angular momentum per unit mass and will form an (energetically favored)  disk configuration as it spirals in~\cite{LyndenBell:1969yx}.
  Because of effective viscous processes (probably due to magneto-hydrodynamic turbulence~\cite{Balbus1991,Balbus1998}), angular momentum is trasferred outwards
and as a result a net mass inflow arises toward the BH.
The gravitational energy of the gas will be dissipated into heat,
which will be either radiated away or advected directly into the BH.

Accretion disk models can be classified according the accretion rate $\dot{M}$ (see e.g. Ref.~\cite{Bambi:2018thh} for a review). A natural scale for accretion is given by the Eddigton accretion rate
$\dot{M}_{\rm Edd}={L_{\rm Edd}}/{\eta}$, where $L_{\rm Edd}$ is the Eddington luminosity and $\eta$ the disk's radiative efficiency.
For $\dot{M} \lesssim \dot{M}_{\rm Edd}$, the radiative efficiency is sufficient to remove  the heat, and the result is a cold geometrically thin disk~\cite{Shakura:1972te,NovikovThorne1973,Page:1974he}.
The case $\dot{M} \gtrsim \dot{M}_{\rm Edd}$ corresponds to a thick disk, where high accretion rates produce  high densities that make the gas optically thick and the radiative transport inefficient. This results in a hot
and ``inflated'' disk~\cite{ThickAccDisksA,ThickAccDisksB}. If instead $\dot{M} \ll  \dot{M}_{\rm Edd}$, the radiative transport is not sufficiently effective to cool down the (low density) gas, which therefore expands into
 quasi-spherical configurations, often  referred to as Advection Dominated Accretion Flows (ADAFs)~\cite{Narayan:1994xi,NarayanYi1995a,NarayanYi1995b}.
Note that ADAFs, even though dynamically very different, are geometrically somewhat similar to spherically symmetric Bondi accretion flows~\cite{Bondi:1952ni}. The latter correspond to purely radial accretion of matter, and are a  good approximation for compact objects accreting gas with negligible angular momentum from the surrounding interstellar medium (ISM).

A common element to all these accretion models is the increase of the matter density as  the BH is approached, even though the density may be
as low as $n_e\sim 10^{-3}$ cm$^{-3}$ far away from it. Notice, for example, that the Bondi accretion model predicts
that the plasma number density should be enhanced by a factor $v_s(\infty)^{-3}$, where $v_s(\infty)$ is the speed of sound at infinity~\cite{Bondi:1952ni, ShapiroTeukolsky1983}. Therefore, number densities close to BH horizons  are expected to be potentially several orders of magnitude higher than in the surrounding ISM. Similarly, ADAF models in the literature also feature very high electron number densities $n_e\sim 10^{19}\left(M_{\astrosun}/M\right)\text{ cm}^{-3}$~\cite{NarayanYi1995b} near the BH horizon.

Ref.~\cite{Conlon:2017hhi} thus concluded that only relatively ``bare'' BHs could be prone to plasma-driven instabilities, e.g. BHs surrounded
by a tenuous plasma because they may have been kicked out of their dense accretion disk after a merger, or because they formed from a violent supernova explosion that blew away most of the stellar material.
Nevertheless, it is not at all clear whether superradiance will occur even under these favorable conditions,
because the increase of the plasma density near the BH (and particularly inside the ergoregion) was not studied by Refs.~\cite{Conlon:2017hhi,pani_loeb}, and may suppress the instability. In fact, it seems likely that the plasma density near the horizon
may increase not only because of accretion, but also because of pair production~\cite{GJ} due to the
large electromagnetic field produced by the instability.

Note  that on physical grounds one would expect the plasma density in the ergoregion, and not at spatial infinity, to play a role,
since the existence of an ergoregion is crucial for superradiance and the Penrose process (as it allows for the presence of the negative energy modes responsible for the extraction of rotational energy from the BH).
Indeed, a well-known semi-analytic result by Eardley and Zouros~\cite{Zouros:1979iw}, valid for scalar perturbations with constant mass $\mu\gg 1/M$ and based on a WKB approximation scheme near the peak of the effective potential,
 seems to suggest that high densities close to the BH would produce instabilities with very long and practically unobservable timescales, i.e. $\tau_I \simeq 10^7 e^{1.84 M\mu} M$ for the fastest growing mode~\cite{Zouros:1979iw}.

However, another important analytic result by Detweiler~\cite{Detweiler:1980uk}, valid in the opposite limit $M\mu \ll 1$
and based on matching two asymptotic wave solutions (one valid near the BH and one near spatial infinity), seems
to suggest instead that only the density at large distances from the BH should matter. Indeed, in the matching procedure
of Ref.~\cite{Detweiler:1980uk} the scalar's mass (corresponding to the plasma frequency and thus to the density) only appears in the solution
valid near spatial infinity, and not in the near-BH solution. The resulting instability timescale is~\cite{Detweiler:1980uk} $\tau_I\sim 48 (\mu M)^{-9} M/\chi$.

In the light of the conflicting intuition from these analytic results, we will undertake in this paper a detailed numerical analysis of superradiance
for fast spinning Kerr BHs surrounded by tenous but accreting plasmas.
To this end, we adopt a simplified toy model where  we represent the electromagnetic field propagating in a plasma by a scalar field with
a position-dependent mass. The dependence on position is required to identify the mass with the plasma frequency, whose local value
changes with the density. We evolve the Klein-Gordon equation for this toy scalar field in the time domain, by using
a spectral technique that was introduced in Ref.~\cite{Dolan:2012yt}, and
which allows for efficiently integrating over long  timescales. We consider several choices
for the density profile of the plasma, in order to explore the impact of the different astrophysical accretion models
outlined above.

This paper is organized as follows. In Sec.~\ref{sec:physical_model}, we outline the physical setup and present
several models for the plasma density profile that we will employ in this paper. In Sec.~\ref{sec:TD} we present our numerical method, while
in Sec.~\ref{sec:results} we describe our results. Our main conclusions are discussed in Sec.~\ref{sec:discussion}.
Throughout this work, we will adopt a signature $(-,+,+,+)$ for the metric. Partial time derivatives will be denoted by an overdot, and
radial derivatives by a prime.

\section{Physical Model} \label{sec:physical_model}
\subsection{Background and perturbation equations}
In general relativity, the spacetime of a rotating BH is described by the Kerr vacuum solution of the Einstein field equations. In Boyer-Lindquist coordinates $\{t,r,\theta,\phi\}$, the corresponding line element is
\begin{multline}\label{eq:kerr}
ds^2=g_{\mu\nu}dx^{\mu}dx^{\nu}\\=-\Big(1-\frac{2Mr}{\rho^2}\Big)dt^2-\frac{4aMr\sin^2\theta}{\rho^2}dtd\phi+\frac{\rho^2}{\Delta}dr^2+\rho^2d\theta^2+\Big(r^2+a^2+\frac{2a^2Mr\sin^2\theta}{\rho^2} \Big)\sin^2\theta d\phi^2\,,
\end{multline}
where $M$ is the mass of the BH, $a= \chi M$,
$\Delta=r^2-2Mr+a^2$ and $\rho^2=r^2+a^2\cos^2\theta$.
On this background, we study the evolution of scalar perturbations with a mass term depending on $r$ and $\theta$ (to be specified in detail in the following),
as a toy model for photons propagating in a plasma surrounding the BH.

The evolution of the perturbations is governed by the Klein-Gordon equation on a curved background:
\begin{equation}\label{eq:KG}
\left(\Box-\mu^2(r,\theta)\right)\Psi=0\,.
\end{equation}
The explicit form of the d'Alembertian differential operator is given by
\begin{equation}\label{eq:laplace}
\Box\Psi=\frac{1}{\sqrt{-g}}\partial_{\mu}\left(\sqrt{-g}g^{\mu\nu}\partial_{\nu}\Psi\right)\,,
\end{equation}
where $g$ is the determinant of the metric.

Since the Boyer-Linquist azimuthal coordinate $\phi$ is known to be singular on the Kerr event horizon, we
change it to Kerr-Schild angle, $\varphi$, defined by
\begin{equation}\label{eq:kerrschild}
d\varphi=d\phi+\frac{a}{\Delta}dr.
\end{equation}
We also change the radial coordinate $r$ to the tortoise radial coordinate, $x$, defined by
\begin{equation}\label{eq:tortoise}
dx = \frac{r^2+a^2}{\Delta} dr\,.
\end{equation}
Using then Eqs.~\eqref{eq:kerrschild} and~\eqref{eq:tortoise} in Eq.~\eqref{eq:KG}, we obtain the following explicit expression for the Klein-Gordon equation
in our coordinates:
\begin{multline}\label{eq:nonsep}
 \Big[\Sigma^2 \partial_{tt}+4aMr\partial_{t\varphi}-(r^2+a^2)^2\partial_{xx}-2a(r^2+a^2)\partial_{x\varphi}+2a^2\Delta \partial_{x}+\frac{2a\Delta}{r}\partial_{\varphi} \\
 +\Delta \Big(-\partial_{\theta\theta}+ \cot\theta\partial_{\theta} +\frac{1}{\sin^2\theta}\partial_{\varphi\varphi}\Big)+\Delta\Big(\frac{2M}{r}-\frac{2a}{r^2} +(r^2+a^2\cos^2\theta)\mu^2(r,\theta)\Big)\Big]\Psi=0\,,
\end{multline}
where $\Sigma^2=(r^2+a^2)^2-\Delta a^2 \cos^2\theta$.\\
From Eq.~\eqref{eq:nonsep}, one can observe that the separability of the perturbation equations
in a radial and an angular part depends crucially on the effective mass term.
Indeed, only the special choice $\mu^2(r,\theta)=(\mathcal{F}(\theta)+\mathcal{G}(r))/(r^2+a^2\cos^2\theta)$
renders the equations separable~\cite{Cardoso:2013opa}.
Except for this special case, the equations are non-separable, and the properties of the perturbations (with particular regards to
their spectrum and their possible superradiant instabilities) are more conveniently computed in the time domain (i.e. via
an initial value evolution) than in the frequency domain.


\begin{table}[t]\label{tab:mass}
\begin{center}
\begin{tabular}{c|l}
	  \hline
      \textbf{Model} & \textbf{Mass profile} \\
      \hline
      \hline
      \textbf{(I)}&  $\mu_0^2(r)=\mu_H^2\left(\frac{r_+}{r}\right)^{\lambda}$ \\
      \textbf{(II)}& $\mu_0^2(r)\sin^2\theta$  \\
      \textbf{(III)}&$\mu_0^2(r)+\mu^2_{c}$ \\
      \textbf{(IV)}& $\mu_0^2(r)\sin^2\theta+\mu^2_{c}$ \\
      \textbf{(V)}& $\mu_1^2(r)=\mu_H^2\Theta(r-r_0)\left(1-\frac{r_0}{r}\right)\left(\frac{r_0}{r}\right)^{\lambda}$ \\
      \textbf{(VI)}& $\mu_1^2(r)\sin^2\theta$\\
      \textbf{(VII)}& $\mu_1^2(r)+\mu^2_{c}$\\
      \textbf{(VIII)}& $\mu_1^2(r)\sin^2\theta+\mu^2_{c}$\\
      \hline
    \end{tabular}
\caption{Mass terms considered in this paper. The effective mass at the horizon is chosen in the range $\mu_H=(1$--$5)M^{-1}$, corresponding to $n_H\sim O(10)$ -- $O(10^2)(M_{\astrosun}/M)^2 \text{ cm}^{-3}$ .The constant mass term can take the values $\mu_{c}=\{0.1,0.2,0.3,0.42,0.5\}M^{-1}$, with corresponding densities in the range $n_c\sim O(0.1)$ -- $O(1)(M_{\astrosun}/M)^2 \text{ cm}^{-3}$. The slope $\lambda$
is chosen among $\lambda=\{1/2,1,3/2,2\}$. For models featuring an inner edge, the latter is placed at $r_0=\{r_{ISCO},3,6,8\}M$.}
\end{center}
\end{table}

\subsection{Mass terms}
The various mass terms that we consider (corresponding to
different density profiles for the plasma) are summarized in Table~\ref{tab:mass}.
Model \textbf{(I)} aims to qualitatively describe Bondi spherically symmetric accretion. The latter predicts a power-law density profile~\cite{Bondi:1952ni},
which in turn gives, through Eq.~\eqref{eq:disprel}, a mass term
\begin{equation}\label{eq:bondi}
\mu^2_0= \mu_H^2\left(\frac{r_+}{r}\right)^{\lambda}.
\end{equation}
The normalization is provided by the mass $\mu_H$ at the horizon $r_+$, while the radial profile is set by the slope $\lambda$. In this paper, we explore values
$\mu_H=(1$--$5)M^{-1}$,
which can be converted [via Eqs.~\eqref{eq:disprel} and \eqref{eq:masscoup}] into
plasma densities near the horizon $n_H\sim O(10)\text{ -- }O(10^2)(M_{\astrosun}/M)^2\text{ cm}^{-3}$.

We adopt such low values of the density to focus on the case of BHs radially accreting from the ISM, like in Ref.~\cite{Conlon:2017hhi}.
As we will discuss in Section \ref{sec:results},
larger values of $\mu_H$ will not produce superradiant instabilities.
Bondi accretion in the transonic flow regime would predict a slope $\lambda=3/2$, but we also explore the impact
of different values $\lambda=\{1/2,1,3/2,2\}$.

In model \textbf{(II)}, we multiply the mass term of model  \textbf{(I)} by  $\sin^2\theta$:
\begin{equation}
\mu^2(r,\theta)= \mu_0^2(r)\sin^2\theta\,.
\end{equation}
 Model \textbf{(II)} therefore attempts to capture the effect of  an axisymmetric ``thick'' disk that qualitatively realizes the ADAF models mentioned in the introduction.
The case of a much thinner disk than model \textbf{(II)} is difficult to study with our code, for reasons that
we will discuss in the following. Nevertheless, we will make the case that model \textbf{(II)}  captures the main qualitative effect
of axisymmetric accretion.

In order to understand the interplay between the values of the density (and effective mass) far away from and close to the BH,
in models \textbf{(III)} and \textbf{(IV)} we consider respectively the mass terms
\begin{equation}
\mu^2(r)=\mu_H^2\left(\frac{r_+}{r}\right)^{\lambda}+\mu_{c}^2
\end{equation}
and
\begin{equation}
\mu^2(r,\theta)=\mu_H^2\left(\frac{r_+}{r}\right)^{\lambda}\sin^2\theta+\mu_{c}^2~,
\end{equation}
where the additional constant term serves as a non-trivial asymptotic value $\mu(r\rightarrow\infty)=\mu_c$, and we choose $\mu_{c}=\{0.1,0.2,0.3,0.42,0.5\}M^{-1}$
[corresponding to plasma densities $n\simeq \{0.1,0.5,1.2,2.3,3.2\}(M_{\astrosun}/M)^2\text{cm}^{-3}$]. We recall that $\mu=0.42M^{-1}$, in the \textit{constant mass} case,
gives the fastest growing superradiant mode for $a=0.99M$~\cite{Dolan:2012yt}, which will be also our choice for the spin parameter.

In order to account for the possibility that the accretion disk may be truncated at some finite distance from the BH, we also
consider the effective mass radial profile
\begin{equation}\label{eq:edgeprof}
\mu_1^2(r)= \Theta(r-r_0)\mu_H^2\left(1-\frac{r_0}{r}\right)\left(\frac{r_0}{r}\right)^{\lambda}\,,
\end{equation}
where $r_0$ is the radius of the disk's inner edge. In our numerical experiments, we choose
$r_0=\{r_{ISCO}, 3, 6, 8\}M$, where $r_{ISCO}$ is the radius of the innermost stable circular orbit around a Kerr BH.
This radial profile is employed, respectively with and without a $\sin^2\theta$ factor, in models \textbf{(V)} and \textbf{(VI)}.

Finally, models \textbf{(VII)} and \textbf{(VIII)}
only differ from models \textbf{(V)} and \textbf{(VI)} because of the addition of a constant mass term $\mu_{c}=\{0.1,0.2,0.3,0.42,0.5\}M^{-1}$. The latter
allows for mimicking the presence of a spherical ``corona'' inside the disk's inner radius,
whose density is non-vanishing but suppressed relative to that of the disk.

\section{Numerical method} \label{sec:TD}
\subsection{Spectral decomposition}
Our time-domain evolution code for scalar perturbations with a space-dependent mass term
utilizes the setup described in Ref.~\cite{Dolan:2012yt} for the constant mass case. We refer the reader to
that work for more details, and we focus here solely on the changes that we had to introduce to deal with a non-constant
mass term.

The method is based on a decomposition of the scalar field in a series of spherical harmonics (see e.g. appendix A in Ref.~\cite{Hughes:2001jr}):
\begin{equation}\label{eq:decomp}
\Psi(t,r,\theta,\phi)=\sum_m\sum_{l=|m|}^{\infty}\frac{\psi_{lm}(t,r)}{r}\mathbf{Y}_{lm}(\theta)e^{im\phi}\,.
\end{equation}
By inserting this decomposition into Eq.~\eqref{eq:nonsep}, we obtain a set of coupled partial
differential equations in the $t$ and $x$ variables. Because of axisymmetry, different \textit{m}-modes decouple from one another, but the decomposition in spherical harmonics generates couplings for each \textit{l}-mode to the $\mathit{l}\pm 2$ modes.

 A first set of couplings arises from the $\cos^2\theta$ terms present both in the coefficients of the time derivatives of the field and in the coefficient in front of the mass term [c.f. Eq.~\eqref{eq:nonsep}]. The projection of these terms on the basis of spherical harmonics can be computed by using
\begin{equation}\label{eq:coscoup}
c^m_{jl}= \langle l m | \cos^2\theta | j m\rangle = \frac{\delta_{lj}}{3}+\frac{2}{3}\sqrt{\frac{2j+1}{2l+1}}\langle j,2,m,0|l,m\rangle \cdot \langle j,2,0,0|l,0\rangle ,
\end{equation}
where we have defined
\begin{equation}\label{eq:proj}
\langle l m | f(\theta) | j m\rangle=2\pi\int_{-1}^{1}\mathbf{Y}^*_{lm}(\theta)f(\theta)\mathbf{Y}_{jm}(\theta)d(\cos\theta)\,,
\end{equation}
and the notation $\langle j_1,j_2,m_1,m_2|j_3,m_3\rangle$ is used for the Clebsch-Gordan coefficients~\cite{AngMom}.

The $\cos^2\theta$ terms generate couplings to $\pi_{l\pm 2}$, which are present also in the massless case, and to $\psi_{l\pm 2}$, which appear in the constant mass case. Both of these  ``classes'' of couplings are  ``geometric'' in nature,
as they arise from the $g^{tt}$ element of the inverse metric. Let us stress that both classes of couplings can in principle be eliminated by projecting onto a basis of
spheroidal (rather than spherical) harmonics~\cite{fletcher1959, Press:1973zz},  at least in the constant mass case. Note however that
spheroidal harmonics are not easy to manipulate in practice, since there are no general analytic expressions for them. The latter
is presumably the reason why Ref.~\cite{Dolan:2012yt} used spherical harmonics even in the constant mass case.

Another different set of \textit{l}-couplings arises from the angular dependence of the effective mass term. For this reason, these couplings do not appear in the
evolution of scalar perturbations with a constant mass term studied in Ref.~\cite{Dolan:2012yt}. In our problem, couplings of this kind are encountered only in the case of the $\theta$-dependent mass profile used in models \textbf{(II)},
 \textbf{(IV)}, \textbf{(VI)}, \textbf{(VIII)}, and can be computed by projecting $\sin^2\theta$ and $\sin^2\theta\cos^2\theta$ as follows:
\begin{equation}\label{eq:sincoup}
s^m_{jl}= \langle l m | \sin^2\theta | j m\rangle = \frac{2\delta_{lj}}{3}-\frac{2}{3}\sqrt{\frac{2j+1}{2l+1}}\langle j,2,m,0|l,m\rangle \cdot \langle j,2,0,0|l,0\rangle ,
\end{equation}
\begin{equation}\label{eq:csscoup}
cs^m_{jl}= \langle l m | \sin^2\theta\cos^2\theta | j m\rangle = \frac{c^m_{jl}}{7}+\frac{3\delta_{jl}}{35}-\frac{8}{35}\sqrt{\frac{2j+1}{2l+1}}\langle j,4,m,0|l,m\rangle \cdot \langle j,4,0,0|l,0\rangle .
\end{equation}
The angular dependence of our mass models therefore generates additional couplings to $\psi_{l\pm 2}$ and $\psi_{l\pm 4}$, as a result of the intrinsic non-separability of the scalar perturbation equations.
Therefore, these couplings cannot be eliminated, even if we were to perform a decomposition into spheroidal harmonics.

Finally, we stress that in practice we cut off the decomposition \eqref{eq:decomp} at a maximum angular momentum number, $\mathit{l}_{max}$, which we vary to check the robustness of our results.

\subsection{Perturbation equations}
By inserting the decomposition \eqref{eq:decomp} into the scalar perturbation equation and introducing the auxiliary variable $\pi= \dot{\psi}$,
we can reformulate the problem in first order form. The result is a system of coupled partial differential equations:
\begin{multline}\label{eq:perteq}
\left(\Sigma_{(0)}^2+a^2\Delta c^m_{ll}\right)\dot{\pi}_{l}+a^2\Delta\left(c_{l,l+2}^m\dot{\pi}_{l+2}+c_{l,l-2}^m\dot{\pi}_{l-2}\right)=\\
\left(r^2+a^2\right)^2\psi''_l+\left[2iam\left(r^2+a^2\right)-2a^2\frac{\Delta}{r}\right]\psi'_l-4iamMr\pi_l-V_0-V_l-V_{l\pm 2}-V_{l\pm 4}\,,
\end{multline}
where we defined
\begin{equation}
\Sigma_{(0)}^2= \left(r^2+a^2\right)^2-a^2\Delta \,,
\end{equation}
\begin{equation}\label{eq:V0}
V_0=\Delta \left[ l\left(l+1\right)+\frac{2M}{r}\left(1-\frac{a^2}{Mr}\right)+\frac{2iam}{r} \right]\psi_l\,,
\end{equation}
\begin{equation}\label{eq:Vl}
V_l= \Delta \Big[\mu^2_{i}(r)\left(r^2+a^2c^m_{ll}+r^2s_{ll}^m+a^2cs^m_{ll}\right)+\mu_{c}^2\left(r^2+a^2c_{ll}^m\right)\Big]\psi_l\,,
\end{equation}
\begin{equation}\label{eq:Vl2}
V_{l\pm 2}= \Delta \Big[\mu^2_{i}(r)\left(a^2c^m_{l,l+2}+r^2s^m_{l,l+2}+a^2cs^m_{l,l+2}\right)+\mu_{c}^2a^2c^m_{l,l+2}\Big]\psi_{l+2}+\left[(l+2)\rightarrow(l-2)\right]\,,
\end{equation}
\begin{equation}\label{eq:Vl4}
V_{l\pm 4}= \Delta \mu^2_{i}(r)\left(a^2cs^m_{l,l+4}\right)\psi_{l+4}+\left[(l+4)\rightarrow(l-4)\right] .
\end{equation}

Eq.~\eqref{eq:V0} gives the effective potential for a scalar  in the Kerr spacetime. That potential  is obviously
common to all the mass models that we consider (and to the massless and constant-mass problems as well). In Eqs. \eqref{eq:Vl} and \eqref{eq:Vl2},
the terms proportional to $\mu_c^2$ are also present in the constant-mass case, while the terms proportional to $\mu_i^2$ are typical of the inhomogenoeus-mass problem tackled in this paper.
 The index $i$ selects between the two radial profiles given by Eqs.~\eqref{eq:bondi} and \eqref{eq:edgeprof}. As already discussed,
terms in \eqref{eq:Vl}, \eqref{eq:Vl2} and \eqref{eq:Vl4} that are proportional to $s^m_{jl}$ and $cs^m_{jl}$ are only encountered in models \textbf{(II)}, \textbf{(IV)} , \textbf{(VI)} and \textbf{(VIII)}, which feature an axisymmetric mass term.

\subsection{Evolution scheme}
We evolve numerically Eq.~\eqref{eq:perteq} by the method of lines. We obtain a set of ordinary differential equations
by approximating the spatial derivatives with a fourth-order finite-difference scheme in the interior of a finite uniform grid in the tortoise coordinate $x$. The grid extends typically from $x_H=-300 M$ to $x_{\infty}=600\text{ -- }1000 M$,
 with typical values of the spacing $\Delta x=0.125 M$. In more detail, we employ a symmetric fourth order approximation scheme for the first and second derivatives of the variables at the inner points of the grid:
\begin{equation}
\psi'_i \approx \frac{-\psi_{i+2}+8\psi_{i+1}-8\psi_{i-1}+\psi_{i-2}}{12\Delta x}
\end{equation}
\begin{equation}
\psi''_i\approx \frac{-\psi_{i+2}+16\psi_{i+1}-30\psi_i+16\psi_{i-1}-\psi_{i-2}}{12\Delta x^2}
\end{equation}
where we have defined $\psi'_i=\psi'(x_i= x_H+i\Delta x)$.
In the neighborhood of the endpoints of the spatial grid we resort to a symmetric but lower order, $O(\Delta x^2)$, derivative scheme
\begin{equation}
\psi'_i \approx \frac{\psi_{i+1}-\psi_{i-1}}{2\Delta x}
\end{equation}
\begin{equation}
\psi''_i\approx \frac{\psi_{i+1}-2\psi_i+\psi_{i-1}}{\Delta x^2}
\end{equation}
At the innermost and outermost points of the grid we impose appropriate boundary conditions, discussed in details in the following subsection.
The time evolution of the equations is performed by a fourth-order Runge-Kutta algorithm with a time-step properly chosen to satisfy Courant-Friederich-Lewy bound for numerical instabilities, $\Delta t= \kappa \Delta x$, with $\kappa<1$. Here, we choose $\kappa=0.8$.
\subsection{Boundary conditions}
Boundary conditions (BCs) play a key role in the study of the spectrum of characteristic modes of a system.
Physical wave solutions in the near-horizon region of a Kerr spacetime
must propagate into the event horizon (which corresponds to $r_* \rightarrow -\infty$), i.e.
they must behave as $\psi \propto e^{-i\omega (t+ r_*)}$ when $r_* \rightarrow -\infty$.
This reflects the known fact that the event horizon effectively behaves as a one-way membrane for classical fields~\cite{Teukolsky:1973ha}.
These BCs, which one can equivalently recast as $\dot{\psi}\approx\psi'$ when $r_* \rightarrow -\infty$,
are usually referred to as ``ingoing'' (into the horizon) BCs,
and are the ones typically adopted to  determine e.g. the spectrum of \textit{quasi-normal modes} (QNMs) of massless scalar perturbations~\cite{Berti:2009kk}.
At spatial infinity, $r_*\rightarrow +\infty$, the general solutions to a wave equation
are comprised of both ingoing (i.e. moving into the grid) and outgoing (i.e. moving away from the systems) modes,
i.e. a generic solution will be $\psi \sim A e^{-i\omega (t+r_*)}+ B e^{-i\omega (t-r_*)}$ (with $A$ and $B$ constants).
If the system is isolated, as assumed in the calculation of QNMs~\cite{Berti:2009kk}, it is appropriate to impose outgoing BCs ($\dot{\psi}\approx -\psi'$)
 to eliminate fluxes entering the system from infinity.
However, when considering fields that have a mass $\mu\neq0$, physical  modes with frequency
$|\omega|\lesssim \mu^{1/2}$ are exponentially (Yukawa) suppressed at spatial infinity.
For this reason, when solving for the \textit{quasi-bound states} (QBSs) of massive perturbations,
one typically adopts ``simple zero'' (i.e. reflective) BCs at spatial infinity,
 $\psi(r\rightarrow \infty)=0$, as considered in~\cite{Dolan:2012yt}.

Implementing proper physical boundary conditions in a numerical method is a non-trivial task.
 In our numerical setup, for example, the left and right grid boundaries are placed at finite values of the tortoise coordinate, and imposing any BC on them generates spurious reflections of the scalar perturbations.
  Ingoing/outgoing BCs involve spatial derivatives of the field, which we approximate with an asymmetric fourth order scheme, e.g. at the innermost point of the spatial grid ($i=0$) we impose
  \begin{equation}\label{inBC}
  \dot{\psi}_{i=0}=\psi'_{i=0}\approx -\frac{11 \psi_{i=0} - 18 \psi_{i=1} + 9 \psi_{i=2} - 2 \psi_{i=3}}{6 \Delta x}~.
  \end{equation}
  Thus, imposing such BCs generates a spurious reflected flux of the same order of the numerical error,
   $\sim O(\Delta x^4)$. For outgoing BCs at the  right boundary, we adopt  the same scheme.
The simple zero BCs behave instead as a perfect mirror: they reflect back the entire incident flux (including the non-superradiant modes) and give rise to unphysical instabilities known as
``BH bombs''~\cite{Cardoso:2004nk}, which could potentially pollute the spectrum of the superradiant modes.

To deal with these artificial reflected scalar fluxes at the left boundary of the grid (i.e. at the horizon), we utilize the same solution suggested in Ref.~\cite{Dolan:2012yt}.
We adopt the finite-difference implementation of an ingoing BC [Eq.~\eqref{inBC}], and we also define a near-horizon region where the equations are modified by the introduction of an artificial damping,
 in the spirit of the \textit{perfectly-matched layers} (PML) technique. This way, the propagation of any spurious reflected signal is effectively suppressed.
  For further details about the PML technique, we refer the interested reader to  Ref.~\cite{Dolan:2012yt} and references therein.

At the right boundary, instead, we observed that for our problem the choice of an outgoing BC is preferable over the simple zero BC used in Ref.~\cite{Dolan:2012yt}.
In fact, we found that an outgoing BC yields smaller reflected scalar fluxes than the simple-zero condition. The reason
of the poorer performance of the simple-zero BC relative to what was observed in  Ref.~\cite{Dolan:2012yt} is probably to be ascribed to
the non-constant mass term that appears in our problem. Since the plasma density, and thus the mass term, increase
when approaching the BH, the potential barrier is higher near the ergoregion in our scenario. A higher potential barrier
is more effective at reflecting back incident modes generated by the spurious reflection at spatial infinity. As a result, these modes remain
trapped between the outer grid boundary and the potential's peak, polluting the numerical evolution. This behavior is instead suppressed if we use outgoing BCs at spatial infinity.

\subsection{Validation}
\begin{figure}[h]\label{fig:dtconv}
\centering
\includegraphics[scale=0.5]{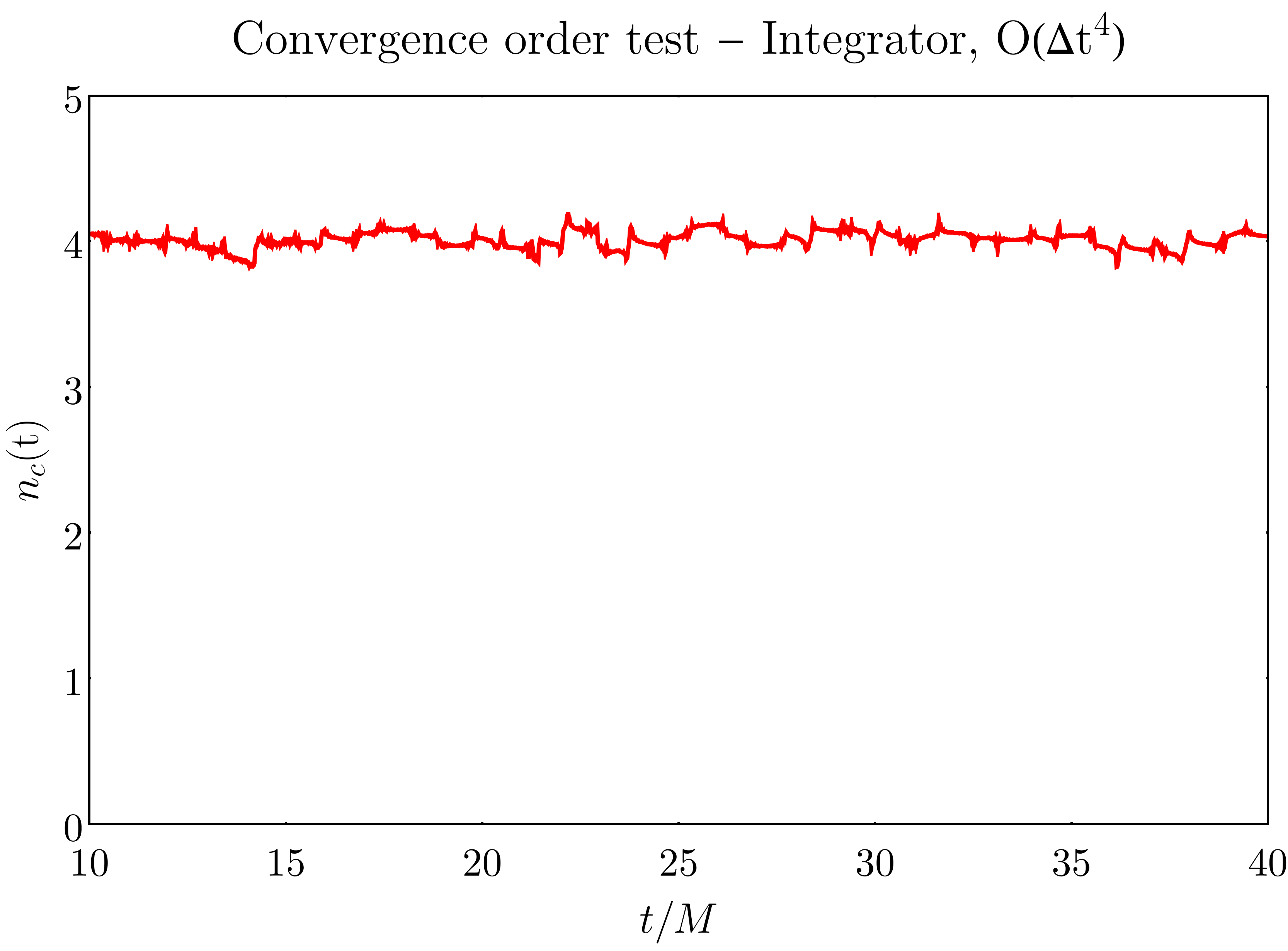}
\caption{Convergence order in $\Delta t$ vs time. The convergence order is estimated as $n_c(t)=\log_2\left(|\Phi_1-\Phi_2|/|\Phi_2-\Phi_3|\right)$, where $\Phi_r \equiv \left(\sum_{l=1}^{lmax} |\psi_{lm;r}(x=0;t)|^2\right)^{1/2}$, with $r$ labelling the resolutions $\Delta t/M=\{0.08,0.04,0.02\}$. The figure shows the moving average of $n_c(t)$ with period $\simeq 1.5 M$.
}
\end{figure}

\begin{figure}[h]\label{fig:dxconv}
\centering
\includegraphics[scale=0.5]{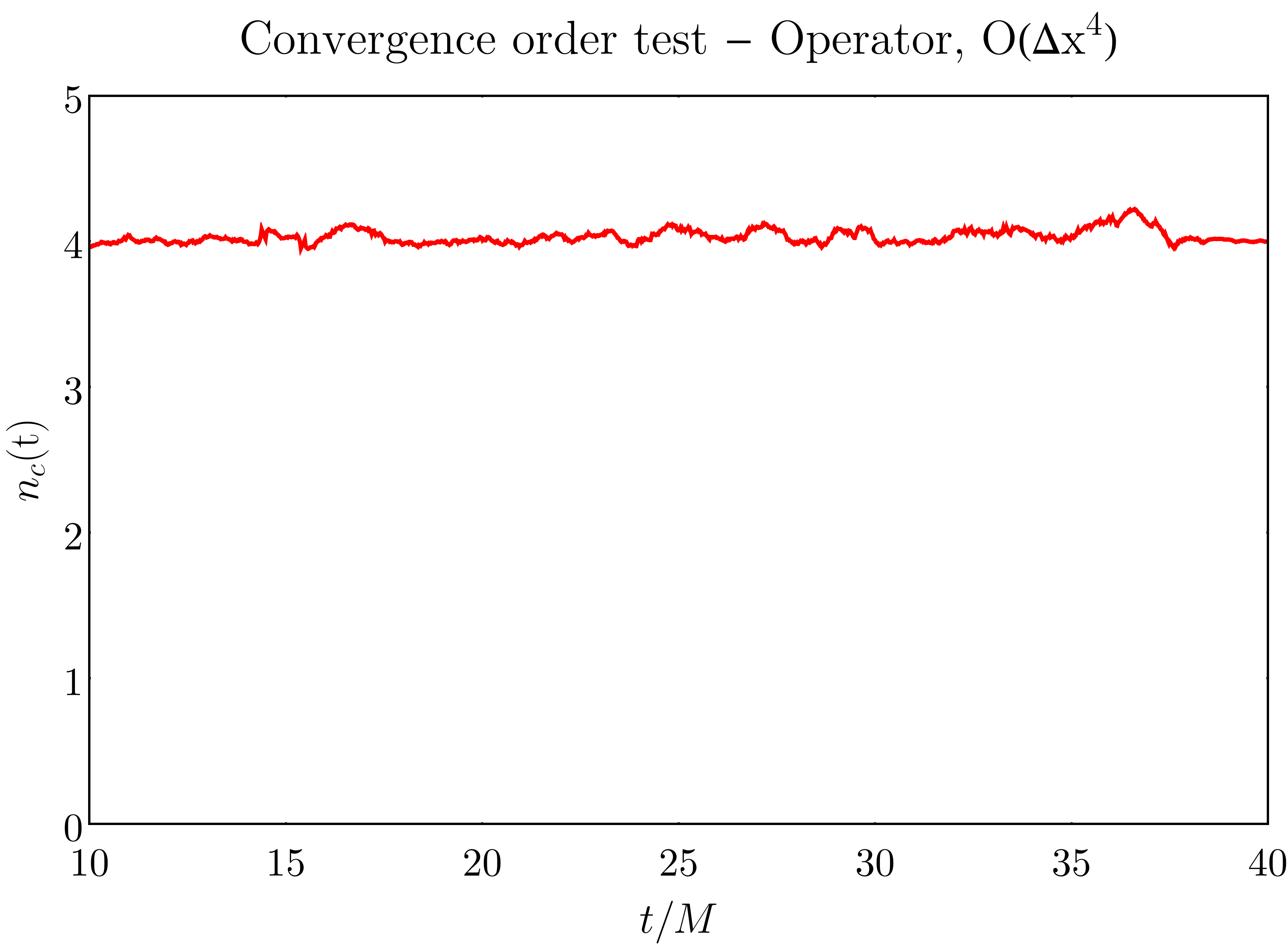}
\caption{The same as Fig.~\ref{fig:dtconv}, but for the convergence order in $\Delta x$ vs time.
The resolutions used are $\Delta x/M=\{0.25,0.125,0.0625\}$.}
\end{figure}

We have performed several tests to validate our results. First, we have tested
that the difference of the results obtained with various time and space resolutions scales
as expected from our finite difference scheme, as shown in Figs.~\ref{fig:dtconv} and~\ref{fig:dxconv}.
 Second,
we have extracted from our evolutions the quasi-normal modes (QNMs) of the scalar perturbations of the Kerr spacetime,
and obtained results in good agreement with the frequencies tabulated in the literature~\cite{Berti:2009kk}.
We have  also computed the superradiant spectrum
for a scalar field with a mirror (BH-bomb) and for a scalar field with a constant mass term,
and found good agreement respectively with the approximated formulae of Ref.~\cite{Cardoso:2004nk} and with
the numerical results of Ref.~\cite{Dolan:2012yt}.
Moreover, we have reproduced the frequency domain results obtained by Ref.~\cite{Cardoso:2013opa}
for a scalar field with a specific mass term yielding separable perturbation equations.
Finally, we have verified that the
total energy and angular momentum of the scalar field (supplemented by the scalar fluxes at infinity and through the horizon) are conserved
to within a good approximation along our numerical evolutions, and we have checked
the robustness of our results against changes of the ``internal'' parameters of our code (e.g. grid size, step, angular momentum cutoff and PML parameters).

\section{Results}\label{sec:results}
In the following, we present examples of numerical results
for the time-domain evolution of the scalar field around a Kerr BH with spin $a=0.99M$, with
the various mass terms reviewed in Sec. \ref{sec:physical_model} and Table~\ref{tab:mass}.

\subsection{Models (I) and (II)}
\begin{figure}[ht]\label{fig:modI}
\centering
\includegraphics[scale=0.8]{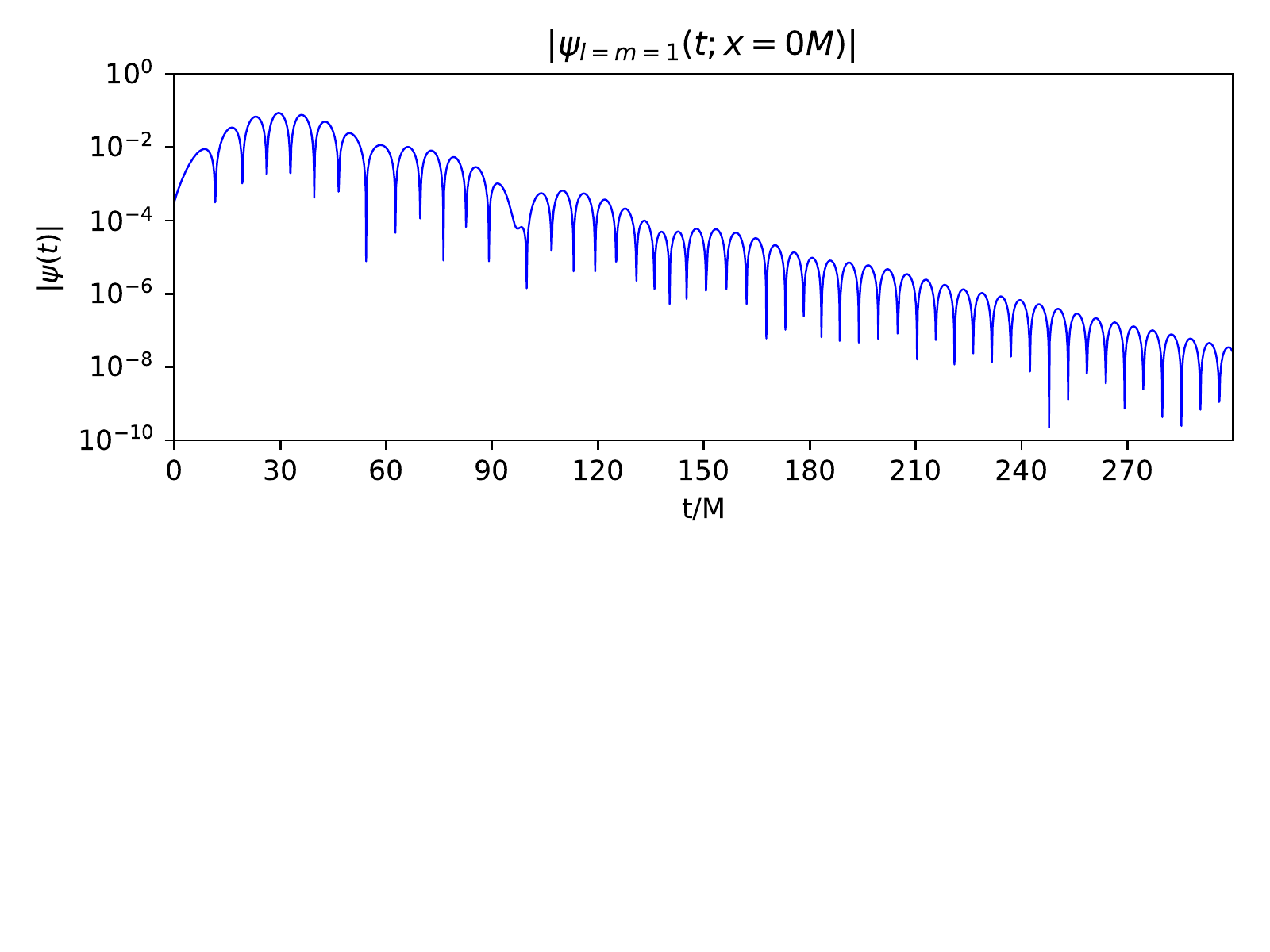}\vspace{-2.7cm}
\caption{Example of QNM ringdown in model \textbf{(I)}, with $M\mu_H=1$ and $\lambda=3/2$: the plot shows the amplitude of $\psi_{lm}$ (with $l=m=1$), extracted at $x=0$.  This mode decays quickly over time, which signals stability. As discussed in the main text,
this is to be expected since mass profile goes to zero at infinity ($\mu(r\rightarrow\infty)=0$).
The extracted QNM frequency  $ M\omega=0.566-i~4.99 \cdot 10^{-2} $ is to be compared with the frequency
for a massless scalar field in Kerr, $ M\omega=0.493-i~3.67\cdot 10^{-2}$~\cite{Berti:2009kk}.
Qualitatively similar results are obtained for the other modes, and for model (\textbf{II}).
}
\end{figure}
We find no evidence for QBSs and for superradiant instabilities in models \textbf{(I)} and \textbf{(II)}, in which the asymptotic mass value at infinity is zero. In fact, in these numerical experiments
the scalar field decays exponentially in time, and the extracted spectrum resembles
the usual Kerr QNM ringdown, though with modified frequencies. A representative example of the spectra that
we obtain is given in Fig.~\ref{fig:modI}, where we show the time evolution of the amplitude of the scalar mode $l=m=1$ in a realization of model \textbf{(I)}.

 From these results, we conclude that a non-vanishing asymptotic mass value at infinity is a necessary condition for the existence of QBSs and
superradiant instabilities. This can be understood by looking at the effective potential  in the limit $a\rightarrow 0$:
if $V_{\rm eff}(r\rightarrow\infty)\rightarrow$ const and $V_{\rm eff}'(r\rightarrow \infty) \rightarrow 0^+$, then the potential features a trapping well that can host QBSs~\cite{Hod:2012zza}. As one can immediately notice, this is not the case for models where $V_{\rm eff}\sim O(1/r^{\lambda})$ at infinity.

As we have already mentioned, however, the QNM frequencies are modified by the presence of a plasma-induced effective mass, with respect to those of a massless scalar on a Kerr background~\cite{Berti:2009kk}. We find that the presence of the plasma can sustain the quasi-normal oscillations for slightly shorter times than in pure vacuum. As expected, in the limit $\mu_H \rightarrow 0$, one recovers the usual Kerr spacetime QNMs.

\subsection{Models (III) and (IV)}
For these models, superradiant modes exist with instability timescales typically longer
than in the corresponding constant mass problem. In more detail, we find instability timescales
 of the order of $\tau_I= \IIm(\omega M)^{-1} \sim 10^{11}M$, which is still shorter than
the typical accretion timescale, and thus potentially relevant in astrophysics.

Nevertheless, these instabilities appear to be very ``fragile'', as they are present only in a small region of the parameter space of models (\textbf{III})
and (\textbf{IV}). In fact, we find no superradiant modes for $\mu_H\gtrsim 2 M^{-1}$, i.e.
a small increase (from $n_H\simeq 0.1 \text{ cm}^{-3}$ to $n_H\simeq 0.5\text{ cm}^{-3}$)  in the density at the horizon is sufficient to quench completely the superradiant instability.
When that is the case, the time evolutions of the scalar field show a damped QNM ringdown, like for models  \textbf{(I)} and \textbf{(II)}, but with typically
longer decay times $\sim 10^{5} \text{ -- }10^{10} M$.
Similarly, as discussed in the previous section about models \textbf{(I)} and \textbf{(II)},
a non-zero value for the mass $\mu_c$ at spatial infinity is needed to get superradiant instabilities,
but as soon as $\mu_cM$ is above a critical value
$\mu_{\rm crit}M \simeq 0.5$ the instability disappears.

The details of the time evolutions depend also on  the
exponent $\lambda$ that controls the slope of the density (and thus mass) profiles: the smaller $\lambda$, the slower the decrease of the mass
profile toward its asymptotic value, and the shorter the lifetime of the stable modes.
Moreover, we also find that there is a critical exponent, $\lambda_{\rm crit}\simeq 2$, below which no superradiant modes exist at all.

\begin{figure}[hp]\label{fig:III}
  \includegraphics[width=0.7\textwidth,center]{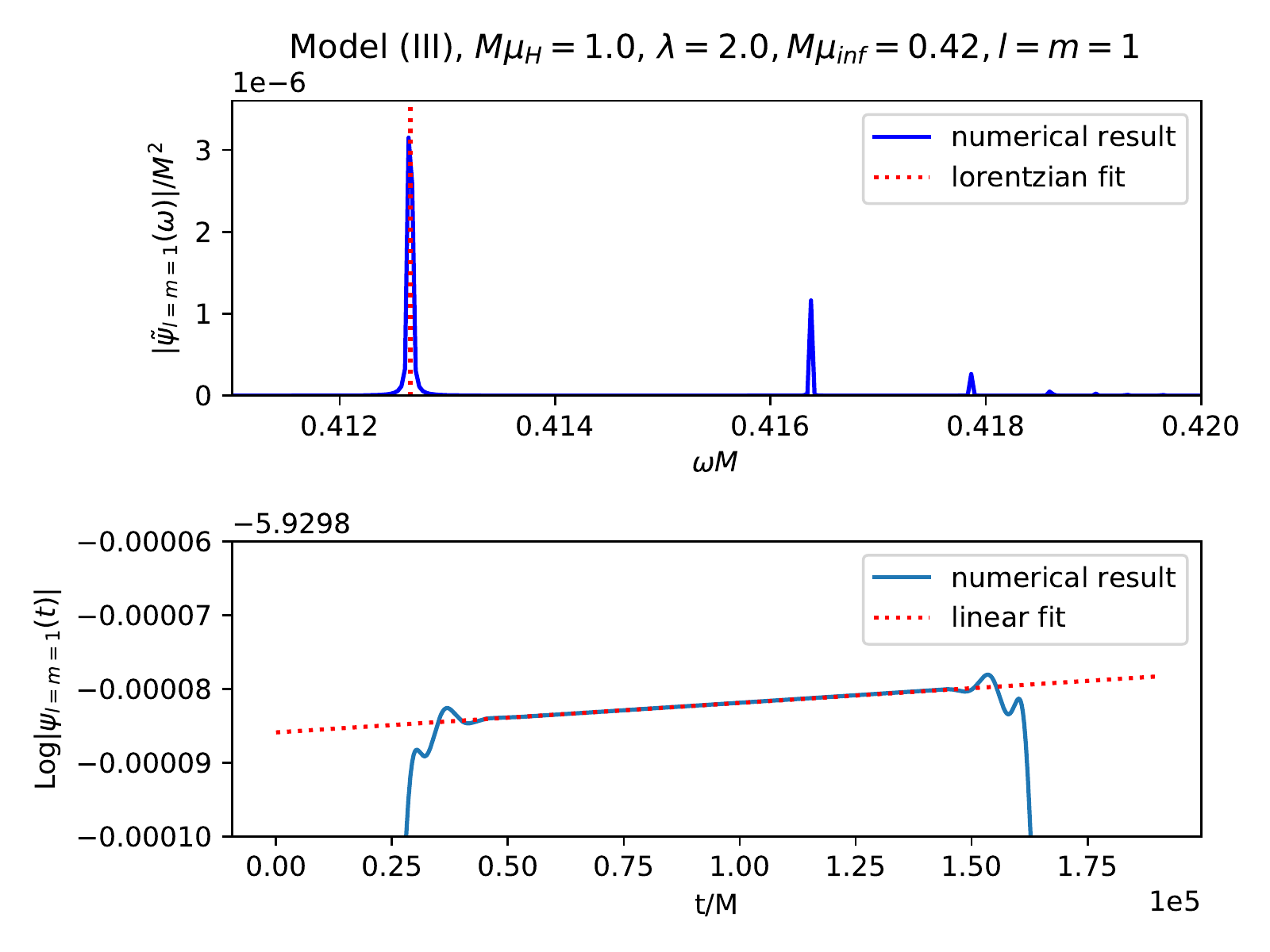}
  \includegraphics[width=0.7\textwidth,center]{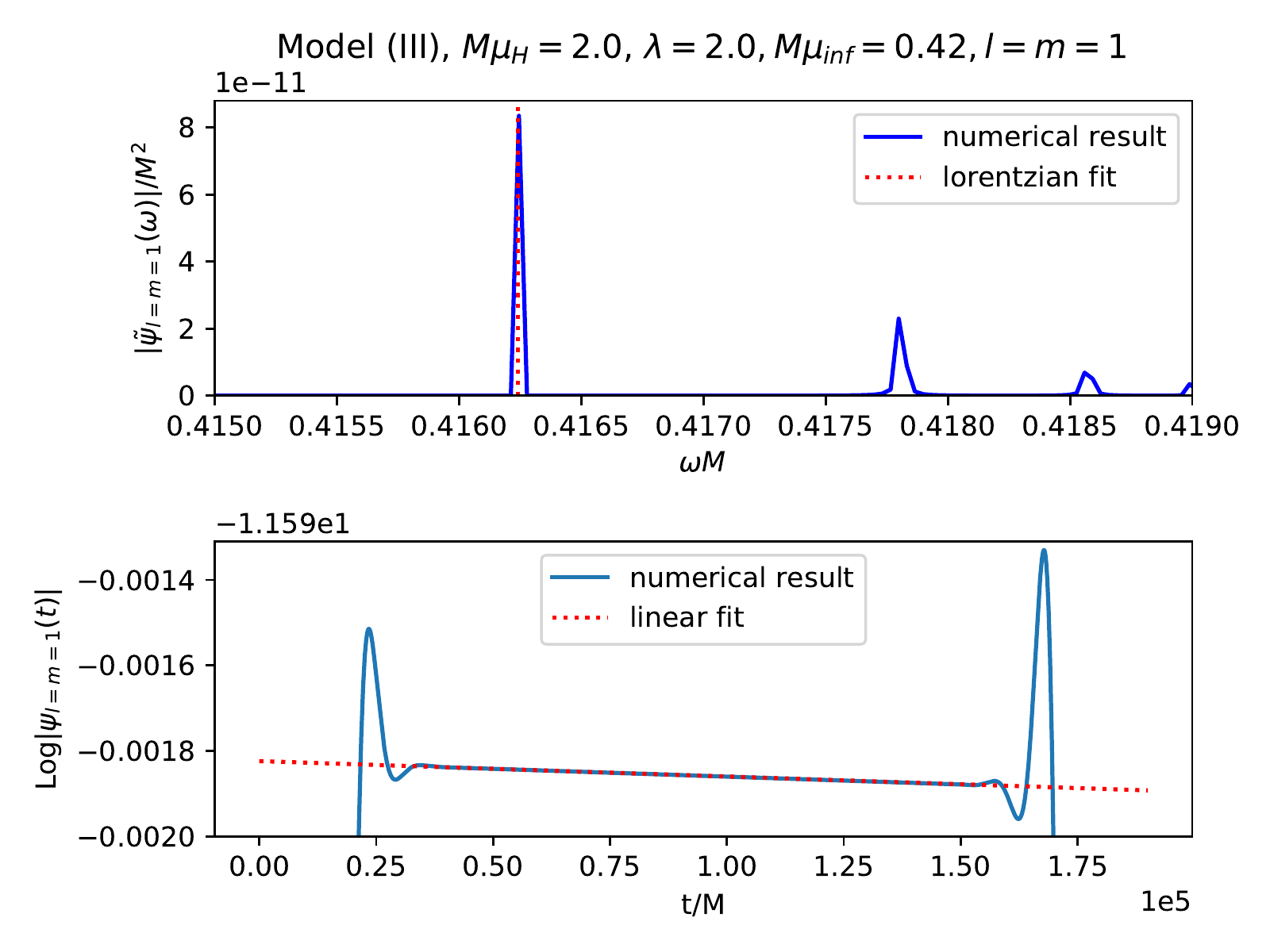}
  \caption{Power spectrum and time evolution for two realizations of model (\textbf{III}), one giving a superradiant instability (top panel) and one giving a stable evolution (bottom panel), for the $l=m=1$ mode.
The two model realizations are respectively one with $M\mu_H=1$ (density at the horizon $n_e\simeq 0.13 \text{ cm}^{-3}$ for a black hole with $M=10 M_{\astrosun}$),
and one with $M\mu_H=2$ ($n_e\simeq 0.52$ cm$^{-3}$ for the same BH mass).
Time evolutions are band-pass filtered to avoid showing the transient due to the initial conditions. This band-pass filter
is responsible for the artifacts at the start and end of the evolution. Dotted lines show Lorentzian fits to the
power spectra and log-linear fits to the time evolutions. The extracted frequencies of the dominant modes are
$M\omega=0.413+i~3.99\cdot 10^{-11}$ and $M\omega=0.416-i~3.61\cdot 10^{-10}$, respectively.
}
\end{figure}

Fig.~\ref{fig:III} shows examples of two scalar field evolutions. The
two upper panels show a realization of model \textbf{(III)} that is subject to superradiant instabilities,
while the two lower panels correspond to  a realization with higher effective mass at the horizon, and thus no instabilities.
In both cases, we show the power spectrum (i.e. the absolute value of the Fourier transform of $\psi_{lm}$, with $l=m=1$),
where one can clearly see the dominant mode and its overtones, as well as a plot of the time evolution, which is
dominated by the main mode and which shows a clear  exponential growth/decay in the unstable/stable case respectively .

Overall, we find that the differences between the spherically symmetric model \textbf{(III)} and the axisymmetric model \textbf{(IV)} are minor, although
``thick'' disks [model \textbf{(IV)}] seem to produce slightly faster instabilities.

\subsection{Models (V) and (VI)}
\begin{figure}[hp]
  \includegraphics[width=0.75\textwidth,center]{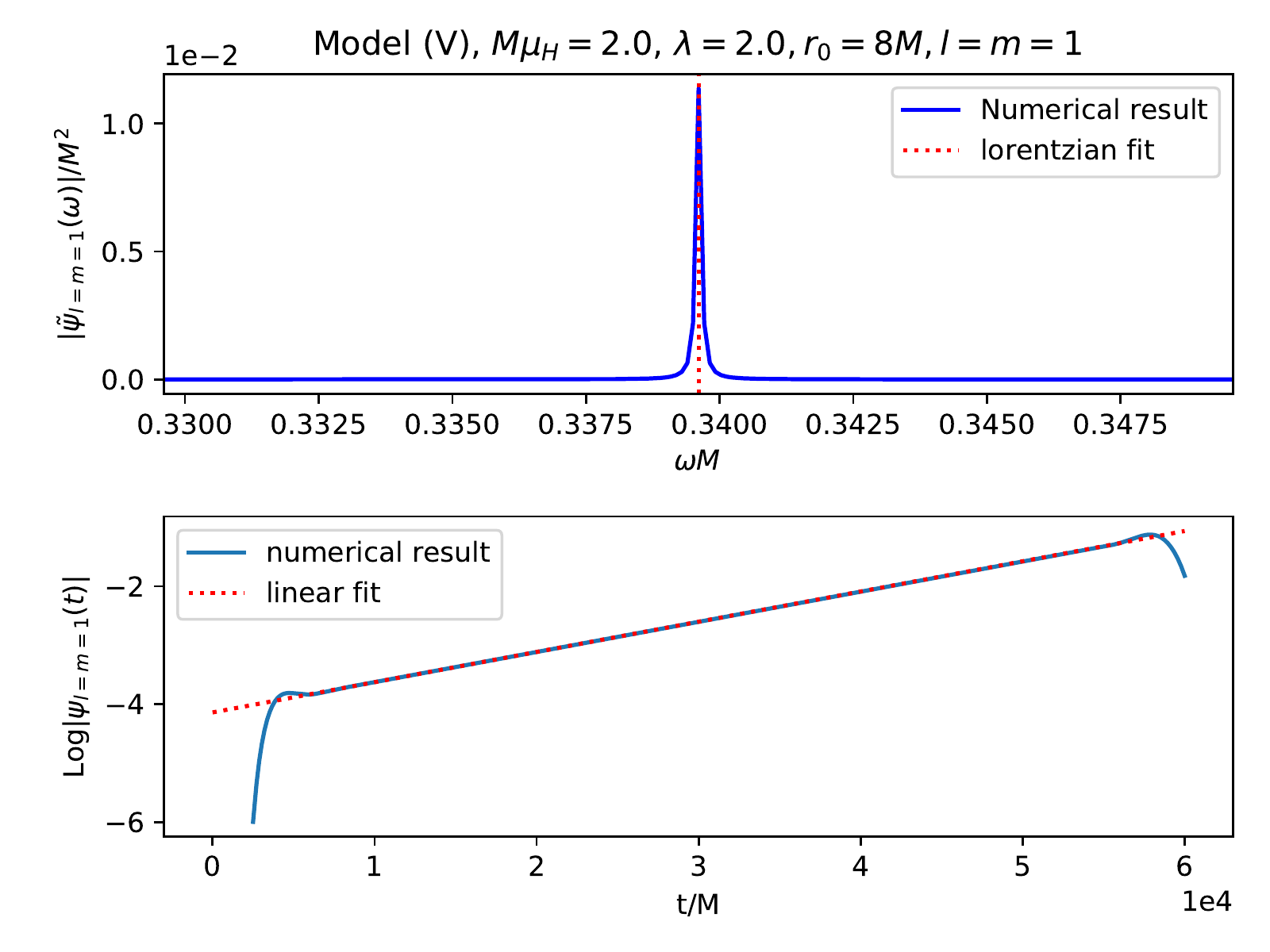}
  \includegraphics[width=0.75\textwidth,center]{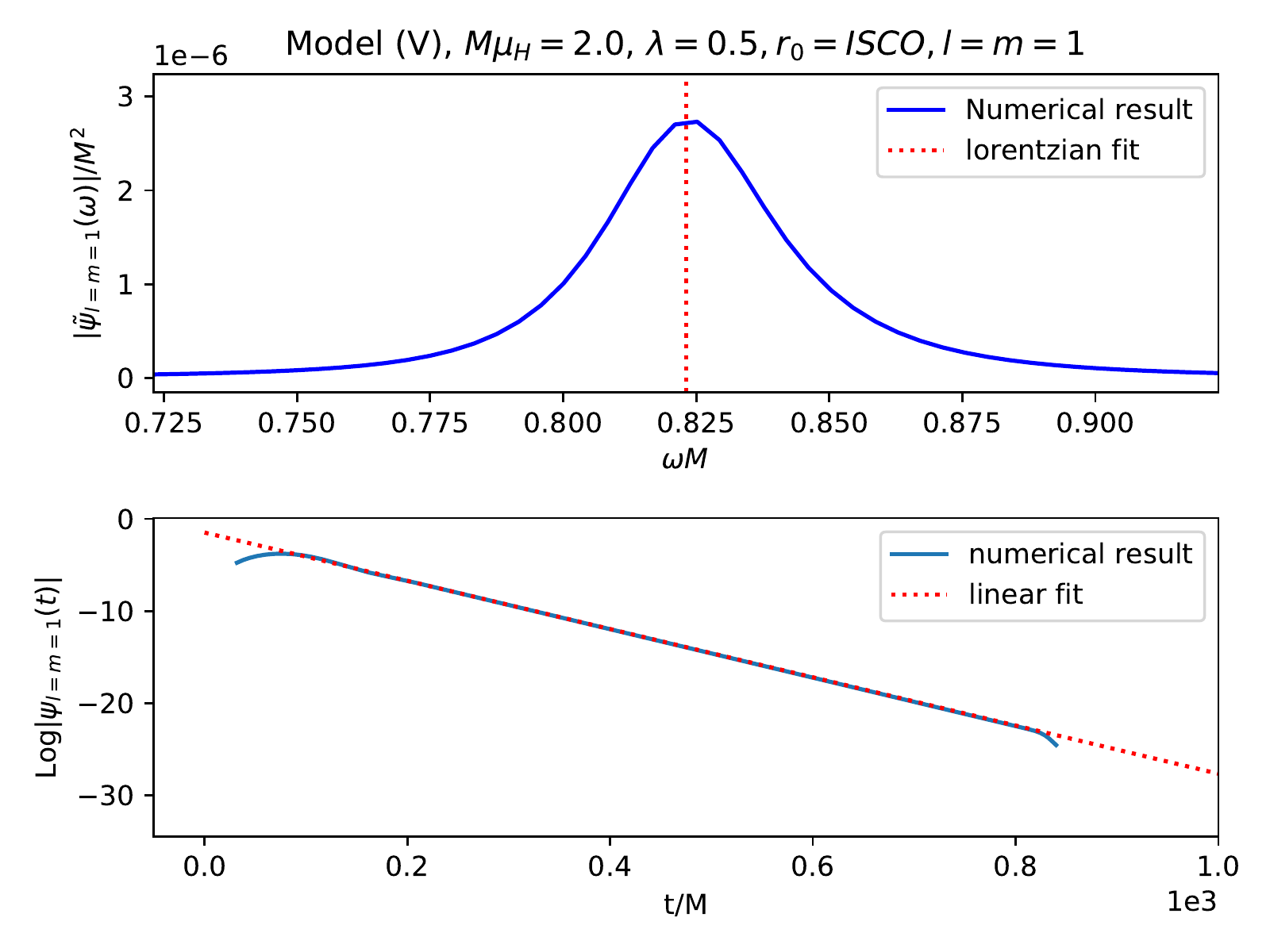}
  \caption{Same as in Fig.~\ref{fig:III}, but for unstable and stable modes ($l=m=1$) obtained with two realizations of model (\textbf{VI}), namely
one with an inner edge at $r_0=8M$ (top panel) and one where density profile is cut off at the ISCO. The top panel shows a  strong instability ($\omega M=0.340+i~5.13\cdot 10^{-5}$), while the bottom one shows a stable evolution with  dominant mode  $\omega M=0.823-i~2.62\cdot 10^{-2}$.
}
\label{fig:V}
\end{figure}

In these models, in which the mass (and plasma density) profiles feature an inner edge but go to zero at spatial infinity, we find both stable QNMs and superradiantly unstable modes. Fig.~\ref{fig:V} shows examples of both.

Mass profiles with the inner edge placed close to the horizon -- $r_0=r_{ISCO}$, $3M$ -- only show signs of stable modes. We have compared the QNM frequencies and decay times extracted from our simulations with the corresponding quantities
 for massless scalar perturbations on a Kerr background~\cite{Berti:2009kk}. We find that, for fixed $r_0$, in the limit $\mu_{H}\rightarrow 0$ one correctly recovers the massless Kerr QNMs: $\RRe(M\omega)\rightarrow 0.493$, $\IIm(M\omega)\rightarrow 3.67 \cdot 10^{-2}$, for $a=0.99M$.
In the opposite limit of increasing $\mu_{H}$, $\RRe(M\omega)$ grows rapidly, while the decay time shows indications of a maximum around $\mu_H\simeq 1 M^{-1}$ and then relaxes to an almost constant value. The latter depends on the choice of the slope and inner edge parameters and is,
 in general, different from the decay time of the QNMs for a massless scalar in Kerr.

Superradiant unstable modes appear instead  only when the inner edge is placed  sufficiently far away from the horizon. To make sense of this result, one can again rely on intuition from
the shape of effective potential in the limit $a\rightarrow 0$. When $r_0\lesssim 3M$, the peak of the mass profile
and that of the effective potential for massless fields are roughly in the same region, which results in a ``flattening'' of
the total effective potential, which in turn prevents the formation of QBSs. For $r_0=6M,8M$, instead, a potential well clearly appears, which
can lead to the formation of QBSs.
In fact, for $r_0=6M,8M$ one obtains fast growing instabilities with $\tau_I \sim 10^5 M$ for spherically symmetric models, while instabilities triggered by ``thick''
disks seem to grow even slightly faster (by a few percent). For these reasons, we expect that similar
(or even stronger) superradiant instabilities should be present even in the limit of razon-thin disks (which we cannot simulate numerically) with inner edge
sufficiently far from the BH.
 Furthermore, the superradiant spectrum resembles closely the results obtained in Ref.~\cite{Cardoso:2013opa}
for  mass terms similar to ours but yielding separable perturbation equations for the scalar.

In spite of these results, we will argue in the following, when dealing with models \textbf{(VII)} and \textbf{(VIII)}, that
in more realistic accretion scenarios the presence of a non-zero (albeit very low) plasma density in a quasi-spherical ``corona'' inside the disk's inner edge
will likely quench these superradiant instabilities.

\subsection{Models (VII) and (VIII)}

\begin{figure}[hp]
  \includegraphics[width=0.8\textwidth,center]{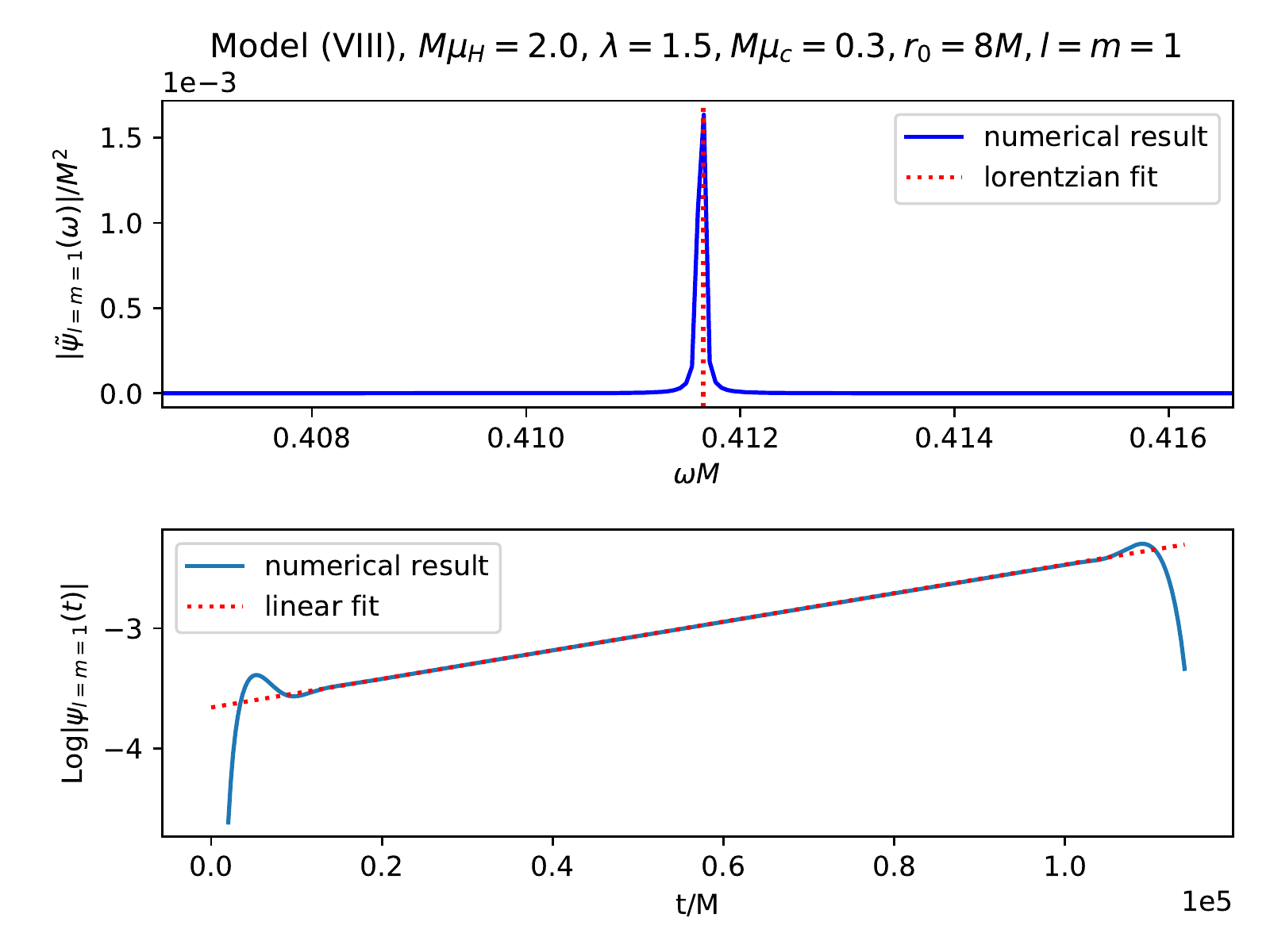}
  \includegraphics[width=0.8\textwidth,center]{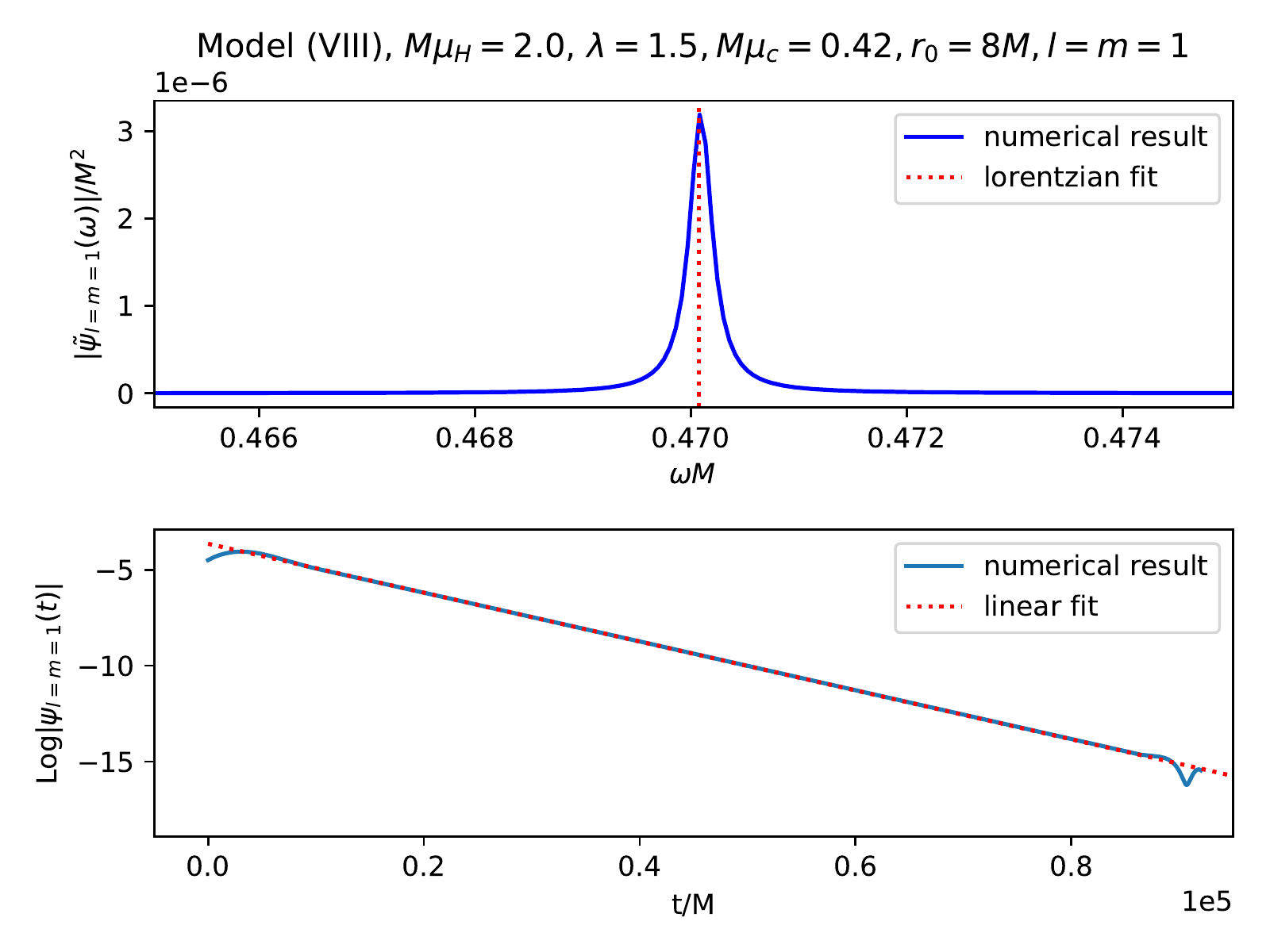}
  \caption{
  Same as in Fig.~\ref{fig:III}, but for two realizations of model (\textbf{VIII}), one with
$M\mu_c=0.3$ (top panels; corresponding to  $n_e\simeq 0.01\text{ cm}^{-3}$ for a BH of $10 M_{\astrosun}$)
and one with $M\mu_c=0.42$ (i.e. $n_e\simeq 0.02\text{ cm}^{-3}$ for a $10 M_{\astrosun}$ BH).
Note that the minor increase of the corona density from the top to the lower panels is enough to quench the instability
($\omega M=0.412+i~1.19\cdot 10^{-5}$ vs $\omega M=0.468-i~1.39\cdot 10^{-4}$ for the top vs bottom case).
Results are for the $l=m=1$ mode.
}
\label{fig:VIII}
\end{figure}

These models show results qualitatively similar to models \textbf{(V)} and \textbf{(VI)}. When the peak
of the mass profile (which is in turn set by the disk's inner edge) is well separated from the centrifugal potential barrier, perturbations can get trapped in a potential well and grow superradiantly with a typical
 timescale of $\tau_I\sim 10^{5}M$.
 Instead, when the peak of the mass profile overlaps with the centrifugal barrier, no QBSs can form and perturbations undergo a
damped ringdown.
\begin{figure}[h!]\label{fig:growth}
 \centering
 \includegraphics[width=.8\textwidth,center]{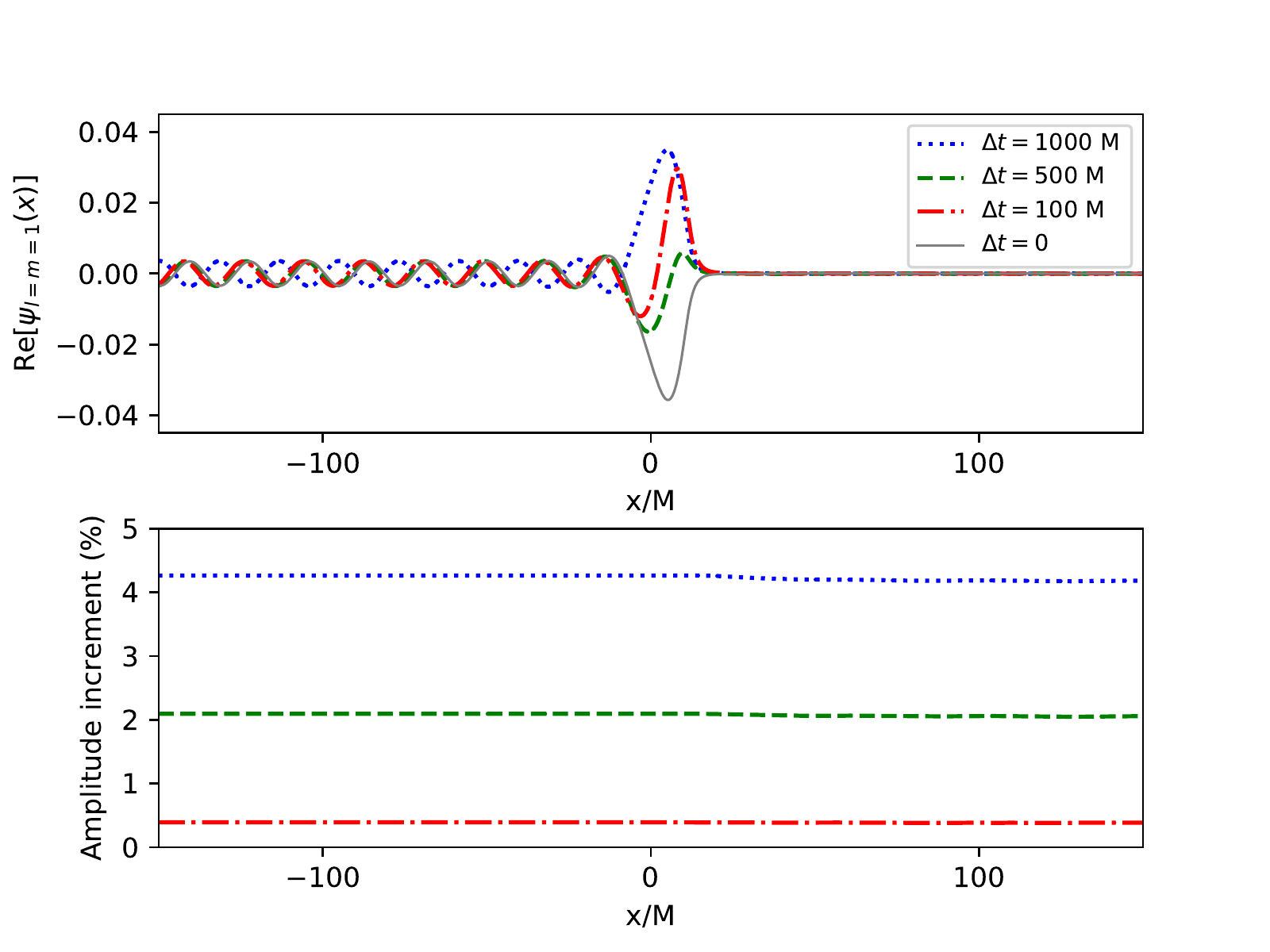}
 \caption{Example of growing superradiant instability (with $\omega M=0.345+i~4.20\cdot 10^{-5}$) in model (\textbf{VIII}), with $\mu_HM=2.0$, $\mu_cM=0.1$, $\lambda=1.5 $, $r_0=8M$.
 The upper panel shows snapshots, taken at different times, of the real part of
 the $l=m=1$ mode over the spatial grid. The lower panel shows the relative fractional amplitude increment relative to some reference time $t=t_0$ (with $\Delta t= t-t_0$).
}
\end{figure}
Fig.~\ref{fig:growth} shows an example of superradiant mode growing over time. In the upper panel, we present snapshots of the quasi-stationary oscillations of a superradiant mode with support in the ergoregion, a
t different times. The lower panel, instead,
shows the fractional amplitude increment of the perturbation over the spatial grid.

Note however that  models \textbf{(VII)} and \textbf{(VIII)}
present a constant mass term $\mu_{c}$ mimicking the presence of a ``corona'', i.e.
a (roughly spherical) region within the disk's inner edge where the accretion flow (and thus the density) are suppressed but non-zero.
Note that astrophysical BHs, and particularly those in intermediate states between ADAF and ``thin'' disk accretion are expected to present this kind of additional structure~\cite{Esin:1997he}.
This corona suppresses  the superradiant modes that were found in models \textbf{(V)} and \textbf{(VI)}.
Indeed, as can be seen from the examples shown in Fig.~\ref{fig:VIII}, the superradiant modes are completely quenched (for both spherically symmetric and axisymmetric models,
 irrespective of the slope $\lambda$) for $M\mu_{c}\gtrsim 0.42$. For a BH of $M=10 M_{\astrosun}$ this corresponds to a very tenuous corona of density $n_e\sim 0.02\text{ cm}^{-3} $. Even higher densities in the corona are expected for realistic accretion scenarios, where the densities in the accretion disk may also be significantly higher. This will have the effect of quenching the instabilities even further,
as larger densities correspond to large scalar field masses, which stabilize the dynamics.
We therefore conclude that realistic accreting BHs are likely safe from superradiant instabilities even when triggered by mass profiles,
such as the ones of models \textbf{(VII)} and \textbf{(VIII)},
that exhibit a sharp cut-off at some inner edge.

\section{Discussion}\label{sec:discussion}
\subsection{Conclusions}
We have investigated the superradiant instabilities that Refs.~\cite{Conlon:2017hhi,pani_loeb} suggested might be triggered
by tenuous plasmas (with densities $n_e\sim 10^{-3}$--$10^{-2}$ cm$^{-3}$ close to those of the ISM) around spinning astrophysical BHs. We have used a 1+1D spectral technique inspired by Ref.~\cite{Dolan:2012yt}
to numerically evolve scalar perturbations with a position dependent mass on a Kerr spacetime.
This scalar is a toy model for the photon field, while the position dependent mass term captures
the effective photon mass  induced by the plasma frequency. The profile of this mass term
is a non-separable function of the radial and polar angle coordinates, and is chosen to mimic astrophysically relevant accretion disk profiles.

From the results of our numerical experiments, we  conclude that a small ($\sim 10^{-3}$--$10^{-2}$ cm$^{-3}$) but non-zero asymptotic plasma density at spatial infinity is crucial for the development of superradiant modes. Indeed, mass (and thus density) profiles that decrease monotonically exactly to zero at spatial infinity do not develop an instability in our simulations.
However, even if the asymptotic plasma density at infinity is small and non-zero,
superradiant instabilities can be easily quenched if the plasma density increases (even slightly) near the BH,
as expected in realistic accretion flows.
This non-trivial interplay between the two asymptotic mass (i.e. density) values, near the horizon and near spatial infinity,
can be qualitatively understood by
looking at the effective potential for the scalar in the limit of vanishing or low spin. Indeed,
one can easily see that while a constant mass term generates a ``trapping well'' where QBSs can form and grow exponentially,
a mass term that increases near the BH does not allow for the formation of mimina (and thus QBSs) in the effective
potential. We find indeed that plasma densities as low as $n_e\sim O(1) (M_{\astrosun}/M)^2\text{ cm}^{-3}$ near the BH
are enough to prevent the the formation of superradiant states.

A notable exception is provided by a plasma density profile exhibiting a sharp cut-off at distances from the horizon larger than a few gravitational radii. If
the plasma density is zero within such an inner edge, superradiant modes can form. However, if the accretion flow (as expected in astrophysically relevant
scenarios) forms a corona with densities as low as $\sim 0.02$ cm$^{-3}$ (for a 10 $M_{\astrosun}$ BH), even these instabilities will be easily quenched.

Overall, our results suggest that astrophysical BHs are likely unaffected by superradiant instabilities.

\subsection{Limitations}
Our work presents several limitations, which we expect should not affect our main conclusions.
First, our numerical integration scheme cannot handle plasma densities that rise too fast as the BH horizon
is approached. Indeed, large plasma densities correspond to large scalar field masses, which make our equations stiff.
As  a  result, in this paper we only consider mass terms as large as $\mu M\sim 5$, which correspond
to  $n_e \sim 3 \text{ cm}^{-3}$  for a BH with mass of $\sim 10 M_{\astrosun}$. While implicit-explicit methods~\cite{Asher1995,Pareschi:2010}
would probably allow for dealing with even larger mass terms, the values that
we consider in this paper are already enough to quench superradiant instabilities, and
on general physical grounds larger masses are anyway expected to stabilize the dynamics even further.

Second, our numerical method cannot handle a razor-thin accretion disk, but only ``thick'' disks. The reason
is that to resolve a thin disk one would need to push our spectral decomposition to
multipole numbers $l\to\infty$. Nevertheless, the scenario envisioned by Ref.~\cite{Conlon:2017hhi}, where BHs are immersed
in a tenous ISM plasma, is expected to produce radiatively inefficient geometrically ``thick''
acccretion flows, which we can study with our code. Moreover, densities and accretion rates in geometrically thin
accretion disks are expected to be much larger than in thick disks, which would make the effective mass
term larger, thus suppressing superradiant instabilities even further.

Obviously, another approximation that may impact our work  is the choice of studying simple toy scalar perturbations instead of
a massive photon (i.e. a Proca field). While superradiant instabilities, when present, are generally stronger for
vector modes than for scalar ones~\cite{Baryakhtar:2017ngi,East:2017ovw,Baumann:2019eav}, the effective potential is very similar for scalars and vectors. Therefore,
we expect
 the qualitative arguments that we give in this paper, and which relate the suppression of superradiant modes
to the shape of the effective potential, should hold even in the vector case.

We also stress  that the dispersion relation given by Eq.~\eqref{eq:disprel} and which provides
the effective mass term for the photon is only valid  for an unmagnetized cold plasma.
This approximation is likely to break down as the temperature of the accretion disk rises close to the BH,
where magnetic fields are also expected to be present.
However, if the dispersion relation given by Eq.~\eqref{eq:disprel} is modified,
it is not even clear if superradiant instabilities would arise in the first place, even under favorable conditions.

Note that a further increase of the plasma density near the BH may occur due to
pair production by the large electromagnetic fields produced by the superradiant instability~\cite{GJ}. While
computing this effect is beyond the scope of this paper, as it would require deriving the structure of the BH magnetosphere produced by the instability (which we cannot do in our scalar toy-problem),
it would actually strengthen our results, since it would lead to an even stronger suppression of superradiant instabilities.

\section{Acknowledgments}
We thank D. Blas, S. Dolan and V. Cardoso for useful conversations on various aspects of the research presented here.
We acknowledge financial support provided under the European Union's H2020 ERC Consolidator Grant
``GRavity from Astrophysical to Microscopic Scales'' grant agreement no. GRAMS-815673.
 This work has also been supported by the European Union's Horizon 2020 research and innovation program under the
Marie Sklodowska-Curie grant agreement No 690904.
A.D. would like to acknowledge networking support by the COST Action CA16104.

\bibliographystyle{apsrev4-2}
\bibliography{masterone}

\end{document}